\documentclass[
12pt,								
oneside,							
headsepline,					
appendixprefix,				
natbib 
]
{article}
\usepackage[english]{babel}
\usepackage{multicol}
\usepackage{graphicx} 	
\usepackage{amsmath}
 \usepackage{epstopdf} 
\usepackage{textcomp}
\usepackage{hyperref}
\usepackage{setspace}
\begin{document}

\title{Planck's radiation law, the light quantum, and the prehistory of indistinguishability in the teaching of quantum mechanics}
\author{Oliver Passon\footnote{University of Wuppertal, School of Mathematics and Natural Sciences, Gau{\ss}str.~20, 42117 Wuppertal, passon@uni-wuppertal.de} and Johannes Grebe-Ellis\footnote{University of Wuppertal, School of Mathematics and Natural Sciences, Gau{\ss}str.~20, 42117 Wuppertal, grebe-ellis@uni-wuppertal.de}}

 \maketitle
\begin{abstract}
Planck's law for black-body radiation marks the origin of quantum theory and is discussed in all introductory (or advanced) courses on this subject. However, the question whether Planck really implied  quantisation is debated among historians of physics. We present a simplified account of this debate which  also  sheds light on the issue of indistinguishability and Einstein's light quantum hypothesis. We suggest that the teaching of quantum mechanics could benefit from including  this material beyond the question of historical accuracy.\footnote{This is an author-created, un-copyedited version of an article accepted for publication in European Journal of Physics. Neither the European Physical Society nor IOP Publishing Ltd is responsible for any errors or omissions in this version of the manuscript or any version derived from it. The Version of Record is available online at \url{doi:10.1088/1361-6404/aa6134} ({\em Eur. J. Phys.} {\bf 38} 035404).}      
\end{abstract}

%
%
%
%
%

\section{Introduction}
December 14, 1900  is usually regarded as the birthday of quantum theory since at that day Max Planck  gave the derivation of his law for black-body radiation at a meeting of the German Physical Society \cite{planck1900b}:
\begin{eqnarray}
u(T,\nu)=\frac{8\pi\nu^2}{c^3}\cdot \frac{h\nu}{\exp{(\frac{h\nu}{kT})}-1}.\label{planck2} 
\end{eqnarray}
Here,  $u(T,\nu)$, describes the spectral energy density of a black-body (i.e., an idealized physical body that absorbs all incident radiation) in thermal equilibrium at temperature $T$ in the frequency interval $[ \nu ,\nu+d\nu ]$, $c$ denotes the velocity of light, $k$ denotes Boltzmann's constant and  $h$ is now called Planck's constant.\footnote{This expression gives the energy per volume and frequency. Often Planck's law is equivalently expressed as energy per unit area, steradian and spectral unit (i.e., frequency or wavelength). As noted by Marr and Wilkin \cite{marrandwilkin} the choice of independent variables affects the shape and even the peak position of the distribution. This may complicate the interpretation by students.}   In the derivation Planck had to introduce an ``energy element", $\epsilon=h\nu$, and most textbooks take this as the origin of quantum theory. 

As noted by Nauenberg \cite{nauenberg2016} textbooks usually do not mention how exactly the ``quantum of energy" was introduced by Planck. The original derivation of Planck is usually not contained in today's texts and e.g. Steven Weinberg remarks ``Planck's derivation is lengthy and not worth repeating here, since its basis is very different from what soon replaced it" (Ref.  \cite{weinberg}, p. 3). A plea for a more historical presentation of Planck's law including the role played by experiment was also made by Dougal many years ago \cite{dougal76}.                     
 
In 1900  neither Planck nor anybody else could possibly anticipate the radical implications of this discovery. Further more it is generally acknowledged that Planck was particularly reluctant to draw radical conclusions from the energy quantum $\epsilon$ and latter tried to bring his findings in conformity with established notions. However, if mentioned at all, this is presented as a debate about the {\em meaning} of the quantum of energy, e.g. whether it signifies something of deep conceptual importance or just a formality and calculation device. 

This view   has been shaken by Thomas S. Kuhn in 1978. His book ``Black-body radiation and the quantum discontinuity" initiated the debate whether Planck implied any kind of discontinuity at all. In his own words:
\begin{quote}  
``My point is not that Planck doubted the reality of quantization or that he regarded it as a formality to be eliminated during the further development of his theory. Rather, I am claiming that the concept of restricted resonator energy played no role in his thought [...]."
(Ref.~\cite{kuhn78}, p. 126)
\end{quote}
Stephen Brush remarks  somewhat sarcastically that Kuhn's interpretation ``is now generally accepted by those historians of physics who have read Planck's 1900 papers" (Ref.~\cite{brush2015}, p. 193). While this claim appears exaggerated (see Darrigol and Gearhart for a balanced discussion of the various strands in this debate \cite{darrigol2001,gearhart2002}) it seems fair to say, that the recent historiography tends to the position that Planck was either indeterminate on this issue or still holding continuity notions at that time. Especially Needell, Darrigol and Badino have elucidated Planck's specific use of statistical mechanics and support Kuhn's claim (or rather object to the standard account) for slightly different reasons \cite{needell80,darrigol92,bumpy}.

This paper will present a simplified account of the ambiguity in Planck's work which in turn triggered the debate on the possible interpretations. Our presentation bypasses many technicalities and makes the topic accessible even to undergraduate students. However, historical accuracy is not our main concern and  our argument does not depend  on the position with which one sides. 
The reason is the following: While the question whether Planck himself implied a discontinuity is surely of historical interest it is of less importance in the teaching of physics. For physics education it is more relevant whether black-body radiation as such provides a clear indication for discontinuity, i.e., whether today's student should accept the claim that this phenomenon alone needs the introduction of a discontinuous energy -- no matter if and why Planck was apparently reluctant to draw this conclusion.  Further more we will argue that this debate can not only be used to explore the historical origin of quantum theory but that it also sheds  light on the issue of indistinguishability and Einstein's light quantum hypothesis. Thus, a deeper understanding  of some fundamental concepts and their interrelation may be gained. 

In Sec.~\ref{bbr} we will briefly review the relevant historical context of the black-body radiation problem and Planck's derivation of his law in which a combinatorial formula figures prominently. Sec.~\ref{ambi} deals with the ambiguous reading of this formula which  lies at the heart of the whole debate. On the reading that energy is quantised the corresponding ``energy quanta" may be compared with the ``light quanta" as introduced by Einstein in 1905. This leads us to the prehistory of  quantum indistinguishability (Sec.~\ref{indis}), since already in 1914 Ehrenfest and Kammerlingh Onnes could show that (in our terminology) ``distinguishable" light quanta can not be reconciled with Planck's law. We will end with some concluding remarks in Sec.~\ref{final}.

\section{The historical context and Planck's derivation of the radiation law\label{bbr}}

Already in 1859 Gustav Kirchhoff argued that a black-body in thermal equilibrium should emit a radiation spectrum which does not depend on its shape or material. Thus there should be a {\em universal function} $u(T,\nu)$, describing its spectral energy density at temperature $T$ in the frequency interval $[ \nu ,\nu+d\nu ]$. On the basis of general thermodynamic considerations Wilhelm Wien showed in 1893 that this function of two variables should have the form $u=\nu^3f(\nu/T)$ (``Wien's displacement law"), i.e., the ``radiation problem" was transformed into the determination of a universal function $f$ of one variable only. Wien even suggested a function which satisfies this requirement, the famous ``Wien radiation law": 
\begin{eqnarray}
u(T,\nu)=\frac{8\pi\nu^3}{c^3}b\cdot \exp{-\frac{a\nu}{T}}. \label{wien}
\end{eqnarray}
The two free parameters could be adjusted such that Equ.~\ref{wien}  described the available data until the turn of the year 1899/1900 when new measurements explored the low frequency part more exactly. At that time Planck had been working on the radiation problem already for some years. His starting point was to model the black-body as a set of charged oscillators. Based on electrodynamics  he could show that the spectral energy density relates to the  temporal average of the oscillator energy ${E}(T,\nu)$ (here $\nu$ is the resonance frequency) according to:
\begin{eqnarray} 
u(T,\nu)=\frac{8\pi\nu^2}{c^3} \cdot {E}(T,\nu). \label{p0} 
\end{eqnarray}
Importantly, this relation (derived e.g. in Ref.~\cite{longair2013}, pp. 32ff) does not contain the mass, charge or damping factor of the hypothetical oscillators. In order to arrive at an expression for ${E}$ Planck's strategy was to determine the entropy of the oscillator, $S(E,\nu)$, and to integrate the thermodynamical relation $dS/dE=1/T$. Following this approach Planck could even derive Wien's radiation law in 1899. However, this was based on a {\em ad hoc} entropy function which he believed then was the only choice compatible with the second law of thermodynamics. With the advent of new data in the long wavelength regime Wien's law failed. In reaction to these data Planck tried to modify the entropy function of the oscillators such that it would give the limit for the long wavelength regime ($u\propto T$) that was suggested by the new data. In October 1900 he proposed his radiation law \cite{planck1900}: 
 \begin{eqnarray}
u(T,\nu)=\frac{8\pi\nu^2}{c^3}\cdot \frac{b\nu}{\exp{(\frac{a\nu}{T})}-1}.\label{planck1} 
\end{eqnarray}
The parameters $a$ and $b$ are the same as in Equ.~\ref{wien}. For high frequencies this expression approximates the Wien law while for low frequencies the so called Rayleigh-Jeans law follows (for $N=\frac{a}{b}R$):
\begin{eqnarray}
u(T,\nu)=\frac{8\pi\nu^2}{c^3} \cdot \frac{R}{N}T . \label{rayleigh-jeans}
\end{eqnarray}
Here, $N$ denotes Avogadro's constant and $R$ the  (molar or universal) gas constant (recall, that $k=\frac{R}{N}$ holds). The Rayleigh-Jeans radiation law follows from the equipartition theorem of statistical mechanics, i.e., inserting $E=kT$ into  Equ.~\ref{p0}. 

Now, many textbooks suggest that Planck somewhat extrapolated between the Wien and the Rayleigh-Jeans radiation law. Given that Equ.~\ref{rayleigh-jeans} leads to a divergent energy when integrated over all frequencies (called ``ultraviolet catastrophe" by Ehrenfest in 1911) textbooks often present Planck's law as a response to this ``crisis". All this can not be quit correct -- and not only because the Rayleigh-Jeans law was only published in 1905 \cite{jeans05}. The historian of physics Martin J. Klein \cite{klein62} gives compelling arguments for the claim that at that time  no awareness of such a crisis existed and that Planck did not realize the importance of the little note Lord Rayleigh \cite{rayleigh1900} had submitted to the Philosophical Magazine in 1900 which anticipated Equ.~\ref{rayleigh-jeans}.

Returning to our main concern, the radiation law as presented in October 1900 (Equ.~\ref{planck1}) was only an educated guess and Planck's immidiate ambition was to derive this  experimentally confirmed prediction from physical principles. This derivation followed the strategy already sketched above. He had to provide the entropy function which (via $dS/dE=1/T$) yields the mean energy of the oscillators such that applying the equilibrium condition (Equ.~\ref{p0}) gives the correct distribution. Since he had guessed the correct distribution already he could work backwards and this is suggested by most historians of physics \cite{rosenfeld36}. His task was thus to provide a theoretical justification for a given entropy function. For this purpose he referred to  Boltzmann's work from 1877 and applied the probabilistic notion of entropy as $S=k\log W$ with $W$ the probability of the corresponding state \cite{boltzmann1877}. Thus, Planck essentially had to count the possibilities of sharing the total energy $E$ among $N$ oscillators. However, a finite result can only be obtained if the energy is considered to be no infinitely divisible quantity. For the number of different ways in which $P$ energy elements $\epsilon$ could be distributed over the $N$ ``resonators" (Planck's expression for the oscillators) he offered the following result \cite{planck1900b}:
\begin{eqnarray}
W=\frac{(P+N-1)!}{P! \cdot (N-1)!}. \label{permu}
\end{eqnarray}
Planck just quoted this relation 
but in 1914 (published a year later) Paul Ehrenfest and Heike Kamerlingh Onnes gave a beautiful and simple derivation \cite{ehrenfest15}. These authors symbolized a distribution of $P$ energy quanta over a number of $N$ oscillators as a string of symbols ``$|$" (the boundary of the oscillators)  and ``$\epsilon$" (the energy unit) such as: ``$\epsilon | \epsilon\epsilon|\;|\epsilon$". In this example we have the energy $E=4\epsilon$ distributed among $N=4$ resonators such that the first resonator contains one quanta, the second two, the third zero and the fourth one. Such a string contains $P$ times the energy symbol and $N-1$ times the line symbol. Hence, $(P+N-1)!$ possible permutations exist. However, given that the $P!$ permutations among the energy symbols and the $(N-1)!$ permutations among the line symbols correspond to the {\em same} distribution, they have to be factored out to reach the number of {\em different} strings, i.e., Equ.~\ref{permu} follows.

In the next step Planck dropped the ``-1" in the factorials, used the approximation $N!\approx N^N$ and could calculate the entropy $S=k\log{W}$ which implies the radiation law:
\begin{eqnarray}
u(T,\nu)=\frac{8\pi\nu^2}{c^3}\cdot \frac{\epsilon}{\exp{(\frac{\epsilon}{kT})}-1}.\label{planck-e} 
\end{eqnarray}
To perform the limit $\epsilon\rightarrow 0$ (as Boltzmann did in 1877) would not yield the desired result  (Equ.~\ref{planck1}).  Further more, the spectral energy density (Equ.~\ref{planck-e}) is a function of $(\epsilon/T)$ while Wien's displacement law demands it to be a function of $(\nu/T)$. Hence, the energy elements have to be proportional to the frequency and Planck set:
\begin{eqnarray}
\epsilon=h\nu. \label{it}
\end{eqnarray}
This gave the law the standard appearance quoted in the introduction (Equ.~\ref{planck2}).

\section{On the interpretation of Planck's combinatorics\label{ambi}}
Most textbooks take it for granted that Equ.~\ref{it}  marks the origin of energy quantisation, but this might be an overhasty interpretation of the mathematical procedure. Thomas S. Kuhn \cite{kuhn78} provides strong  evidence that Planck proposed a ``physically structured phase space rather than discontinuous energy levels" (Ref.~\cite{kuhn80}, p. 187). This view is supported by Planck's remark \cite{planck1900b}:
\begin{quote}
``If the ratio [of the total energy $E$ to the energy element $\epsilon$] thus calculated is not an integer, we take for $P$ an integer in the neighborhood."  
\end{quote}
Thus, the $\epsilon$ can also be viewed as describing an {\em interval} of the still continuous energy.\footnote{However, this ``integer-in-the-neighborhood"-remark was never repeated by Planck -- in particular it is not contained in \cite{planck1901} which summarizes \cite{planck1900b} for the {\em Annalen der Physik}. The upholders of the claim that Planck intended a {\em physical} quantization usually quote his remark, that the energy elements were the ``most essential point of the whole calculation" \cite[p. 239]{planck1900b}. As always, a single quote taken out of context can not prove anything.}

While this debate involves highly complex and technical details,  it originates largely from the following simple ambiguity. Planck's combinatorics can be interpreted in two distinct ways (compare Ref.~\cite{darrigol91}, p. 243ff):
\begin{enumerate}
\item  {\bf Discontinuity reading:} Equ.~\ref{permu} gives the number of ways how $P$ energy elements $\epsilon$ can be distributed over $N$ resonators -- and this interpretation is suggested by the derivation of Ehrenfest and Kamerlingh Onnes (compare Sec.~\ref{bbr}). This view suggests that the absorption and emission is discontinuous indeed. 

\item {\bf Continuity reading:} Equ.~\ref{permu} describes the ways to distribute resonators over ``energy cells" (taking care that energy conservation is not violated). According to this view the resonators are placed in energy cells of finite size, however, they can be put anywhere inside this cell. Hence, a continuous emission and absorption is compatible with this view while only the {\em specific} size of Planck's energy cells is mysterious. 
\end{enumerate}
As pointed out by Badino already in Boltzmann's work from 1877 (which was cited by Planck in 1900) {\em both} interpretations are contained \cite[p. 92ff]{bumpy}.
The debate about Planck and the quantum is to a large degree on the issue with which interpretation of the combinatorics he sided. His own writings are ambiguous and in trying to decide this question additional sources need to be considered, like Planck's general research program and his attitude towards statistical mechanics and thermodynamics. Especially his view on Boltzmann's work and the role of probability  was apparently very idiosyncratic and changed over time. However, it has been noted that one need not  presuppose that Planck had a coherent and articulated view on this issue in 1900 or 1901 at all. This suggests, that Planck might have been undecided and neither the discontinuity nor the continuity position should be attributed to him \cite{galison81}. Badino suggests that exactly this ambiguity in Boltzmann's work enabled Planck to maintain a noncommittal stance to the underlying micro-physics \cite[p. 96]{bumpy}. For example it is well documented that Planck remained convinced in the strict validity of the second law of thermodynamics although he applied Boltzmann's methods of statistical mechanics.

While the question whether Planck himself implied a discontinuity is surely of historical interest it is of less importance in the teaching of physics. For physics education it is more relevant whether black-body radiation as such provides a clear indication for discontinuity, i.e., whether today's student should accept the claim that this phenomenon needs the introduction of a discontinuous energy. 


No matter how (and if) one decides on the historical matter, it is highly interesting that such a conservative reading of Planck's law is possible and valid. This helps also to understand why Planck's work was soon acknowledged as experimentally confirmed radiation law while a debate on ``quantisation" did not follow immediately \cite[p. 23f]{jammer}.

\section{On the early history of indistinguishability\label{indis}}
As is well known, the development of quantum theory gained momentum with Einstein's light quantum hypothesis in 1905 \cite{einstein05}. The misrepresentation of this work is legendary. According to the common textbook narrative Einstein promoted Planck's energy quanta to ``light quanta" and explained the photo-electric effect with them. In fact, his work is completely independent from Planck's law and the photo-electric effect is rather mentioned in passing. His starting point for introducing light quanta is Wien's radiation law (Equ.~\ref{wien}), i.e., black-body radiation at high frequencies. Einstein calculated the entropy of the radiation with total energy $E$ for a given Volume $V_0$ and derived the probability for a fluctuation into the sub-volume $V$ to be:
\begin{eqnarray} 
W_{\mathrm{rad}}=(V/V_0)^{E/h\nu}. \label{lq}
\end{eqnarray}
In fact, Einstein did not even use Planck's constant but an equivalent expression. This equation looks like the probability for an ideal gas of $P$ point-like molecules to fluctuate into the sub-volume:
\begin{eqnarray} 
W_{\mathrm{gas}}=(V/V_0)^{P}.
\end{eqnarray}
Einstein concluded that monochromatic radiation in the Wien regime can be viewed as consisting of $P$ localised light quanta with energy $h\nu$. The conclusiveness of Einstein's argument is  debated. While Dorling has taken it to be a ``rigorous deduction" (i.e., even stronger than Einstein thought) Irons suggests the application of unwarranted assumptions with regard to the statistical fluctuations of the radiation \cite{dorling71,irons2004}. See also Norton for a discussion of Einstein's ``miraculous argument" \cite{norton2006}. However, given that the corpuscular properties of light are only established for the Wien regime they can not claim universal validity anyway.      

While Einstein based his original argument for light quanta on Wien's law he discussed the relation to Planck's law a year later in 1906.  Einstein pointed out that  Planck applies contradictory assumption since Equ.~\ref{p0}   treats the oscillator energy as continuous variable, while subsequently the energy $E$ is taken to vary in discrete steps only. Thus, the basis for the whole derivation has been removed. If we follow the above arguments by Kuhn et al. there is  no reason to assume that Planck really implied the discreteness of energy and the alleged contradiction vanishes.   In fact, Kuhn and Darrigol count the use of Equ.~\ref{p0} as additional evidence for Planck not implying a discontinuity (Ref.~\cite{kuhn78}, p. 118 and Ref.~\cite{darrigol88}, p. 55). Einstein was evidently willing to take the energy elements more seriously and assumed that Planck regarded them as real discontinuity likewise. He concluded that an additional assumption is needed in order to justify the equilibrium condition between the radiation and the oscillator energy expressed in Equ.~\ref{p0}, namely the discreteness of the radiation field. Einstein even made the surprising claim that: ``Herr Planck introduced  into physics a new hypothetical element: the hypothesis of light quanta" (Ref.~\cite{einstein06}, p. 196).   

Now, the light quantum hypothesis was rejected by most physicists until A. H. Compton's  discovery of the effect named after him and its theoretical explanation in 1922/23. But starting with Einstein's work on the specific heat in 1907 \cite{einstein07} the notion of quantised energy levels in solids gained acceptance and the Bohr model of the atom (1913)   applied this notion likewise. Given this growing success of the ``quantum theory" the energy quanta were soon viewed as a true physical discontinuity. In our terminology with regard to the two interpretations of  Equ.~\ref{permu}, we could say that the ``discontinuity reading" was adopted.

The above quote indicates that Einstein apparently equated his light quanta with Planck's  energy quanta. While this claim could be later supported for photons as ``indistinguishable particles"  it is wrong with regard to the light quanta as introduced by Einstein in 1905. The issue of indistinguishability in quantum theory has an exciting prehistory which is rarely mentioned in textbooks. Already in 1911 the Polish physicist W{\l}adys{\l}aw (or latinized ``Ladislas") Natanson scrutinized the statistical assumptions underlying   Planck's law and anticipated this concept   \cite{natanson1911}. 
 Natanson discriminated  between the situation where (i) both, the units of energy and the ``receptacles of energy" can be distinguished, (ii) only  the receptacles of energy can be identified (i.e., are distinguishable ), or,  (iii) only the units of energy  are distinguishable. In either case a different combinatorics needs to be applied. Natanson claims that Planck's Equ.~\ref{permu} assumes the scenario (ii), i.e., treats the energy elements as indistinguishable. But he failed to draw a connection to light quanta.

This connection was drawn by Paul Ehrenfest in 1911 but argued more convincingly  in 1914  by Ehrenfest together with Heike Kamerlingh Onnes in the paper already quoted for the simple derivation of Planck's combinatorial formula \cite{ehrenfest15,ehrenfest11}. In the appendix of this paper the authors address the issue that apparently the only difference between Planck's energy quanta and Einstein's light quanta is that the later also exist in empty space. They argue however, that this claim is incorrect since one needs to consider whether the objects distributed are ``quanta,  existing  independently  of  each  other" as in Einstein's argument or ``a purely formal device" as -- according to these authors -- in Planck's argument. 
Assuming this independence for $P$  quanta distributed over $N$ ``cells in space" would yield $N^P$ combinations. Exactly this kind of combinatorics is expressed by Einstein's Equ.~\ref{lq} from 1905 which was derived from Wien's law. Ehrenfest and Kamerlingh Onnes continue by giving a simple numerical example: If $P=3$ quanta are distributed over $N=2$ cells Einstein needs to count $2^3=8$ different distributions while Planck's formula yields only $\frac{4!}{3!1!}=4$, i.e., identifies those combinations which differ only by a permutation.  Ehrenfest and Kamerlingh Onnes conclude that Einstein's {\em independent} quanta lead to Wien's law only while ``Planck's formal device (distribution of $P$ energy elements $\epsilon$ over $N$ resonators) cannot be interpreted in the sense of Einstein's light-quanta" (Ref.~\cite{ehrenfest15}, p. 301).

Numerical examples as the one above are used in many of today's textbooks to illustrate the difference between Maxwell-Boltzmann and Bose-Einstein statistics. Does this imply that   Ehrenfest  and Kamerlingh Onnes  took the energy quanta as indistinguishable particles? While the need for a different statistics is clearly anticipated by this argument these authors apparently rejected the notion of indistinguishability  and this is why they called Planck's energy quanta a ``formal device"   (compare Ref.~\cite{darrigol91}, p. 239). This should not be misinterpreted as if these authors took the energy quanta as a calculation device only since Ehrenfest was intensively involved in the development of the quantum theory at that time. 

Summing up,  this argument shows that Einstein's light quanta from 1905, modeled after the molecules of classical kinetic gas theory, can not be reconciled with Planck's law and differ from the ``photons" which later gained acceptance. Revolutionary as the concept of light quanta was in 1905, in the sense alluded above it was apparently too classical still. 

A similar result (i.e. that distinguishable light quanta can not be reconciled with Planck's law) can be reached  by studying the volume dependence of the entropy of black-body radiation.  Provost and Bracco have applied Einstein's original argument of 1905 based on Wien's approximation to Planck's law and reached the same conclusion \cite{provost}. However, our discussion is much more basic.


Certainly Planck did not assume any kind of indistinguishability either -- not even tacitly or implicitly. Only later derivations of Planck's radiation law moved the discontinuity from Planck's resonators into the radiation field and Satyendra N. Bose's work from 1924 derives it with the help of ``indistinguishable" light quanta which obey a different kind of statistics. To Planck such a view must have been foreign even if one attributes to him the discontinuity reading of Equ.~\ref{permu}, since the discontinuity is with respect to the oscillator energy only.  According to  the continuity  interpretation this question does not even arise. On this view one distributes {\em distinguishable} resonators (the analogue to Boltzmann's molecules) over cells in energy-space. These cells may be viewed as mathematical abstractions and in any event they are no particles. To assume their being {\em indistinguishable} could not conflict with the classical idea of independent particles at all (compare Ref.~\cite{darrigol91}, p. 249). 
 
\section{Summary and concluding remarks\label{final}}
Historians of science agree that ``discoveries" are rarely attributable to a particular moment in time and sometimes not even to single individuals. They are rather extended processes which involve the interaction of several researchers. But it is often possible to single out individuals who have finally pushed this development to a point from which there could be no retreat. According to Kuhn (Ref.~\cite{kuhn78}, p. 369), Planck is an example for this. After his distribution law for black-body radiation the recognition of discontinuity was inevitable, although he was apparently reluctant to draw this conclusion at first. Thus, the debate whether Planck implied a discontinuity or if a more conservative reading is possible is   instructive since it illustrates this point for a discovery of first rank. To make this point needs of course that Planck's original derivation is introduced. Since Planck's law is such a fundamental result it can be derived in many different ways and most of today's textbooks follow a different route here (e.g. Einstein's proof with the help of his $A$ and $B$ coefficients). Also popular are derivations which are based on the counting of modes in a perfectly reflecting cavity. However, Irons has pointed out that such a system lacks any mechanism for the thermalisation of the radiation and that this problem (if addressed at all) has only questionable attempted solutions \cite{irons2005}.


Given that black-body radiation is a state of equilibrium between the radiation and the emitting (and absorbing) body, it opened up two different research lines at the same time. In a confusing and perplexing sequence of events the notion of quantised matter and quantised radiation developed almost simultaneously. To follow these twists and turns would be a very unfavorable teaching strategy and reconstructed systematically it is the common choice to introduce the non-relativistic quantum mechanics first and to deal with relativistic quantum theory and the quantum theory of radiation only later. However, many textbooks (and popular representations) on quantum theory can not resist the temptation to introduce light quanta already in  connection with Einstein's explanation of the photo-electric effect. The argument presented here can help to emphasize that these light quanta should not be confused with our current photon concept. According to current understanding the photon of QED is neither distinguishable nor localizable, i.e., it is no ``fuzzy ball" \cite{scully72}. 

As mentioned above, the light quantum hypothesis gained strong acceptance with Compton's explanation of his X-ray scattering experiments. But this then accepted light quantum (soon to be called ``photon") was a localised and distinguishable particle still, i.e., should not be confused with the current photon concept either. It is certainly a truism that concepts in physics change over time (while their names remain the same). But unfortunately many textbooks introduce Einstein's explanation of the photo-electric effect and Compton's kinematic derivation of the photon wavelength shift as if these explanations fit into the current understanding of quantised radiation \cite{scully72}.  

Ehrenfest's  simple objection against Einstein's light quanta of 1905 as ``independent" (i.e., distinguishable) particles remains valid.  This argument of 1914 also shows that the rejection of the early light quantum was more rational than commonly presented. To include this exciting prehistory of the Bose-Einstein statistics (which should rather be called ``Ehrenfest-Natanson-Bose-Einstein statistics") helps to make this important distinction while the usual teaching tradition introduces  indistinguishability only as a chapter of the advanced many-particle theory and quantum statistics.

 
\section*{Acknowledgments  }
We gratefully acknowledge an illuminating email exchange with Massimiliano Badino. \\[0.5cm]


\end{document}